# Effect of Sb substitution on the Topological Surface States in $Bi_2Se_3$ single crystals: a magneto-transport study

T. R. DEVIDAS[1(a)], E. P. AMALADASS[1], SHILPAM SHARMA[1], AWADHESH MANI[1(a)], R. RAJARAMAN[1], C. S. SUNDAR[1] AND A. BHARATHI[2]

[1]*Materials Science Group, Indira Gandhi Centre for Atomic Research, Kalpakkam - 603102, Tamil Nadu, India*
[2]*UGC-DAE CSR, Kalpakkam Node, Kokilamedu – 603104, Tamil Nadu, India*

**Abstract** – Magneto-transport measurements have been carried out on $Bi_{2-x}Sb_xSe_3$ (x = 0, 0.05, 0.1, 0.3, 0.5) single crystals at 4.2 K temperature in the magnetic field range of -15 T to 15 T. Shubnikov-de Haas (SdH) oscillations of 2D nature were observed in samples with Sb concentration upto x = 0.3. The analyses of SdH oscillations observed in magneto-resistance data using Lifshitz-Kosevich equation reveal a systematic decrease in the Fermi surface area with Sb substitution. The Berry phase obtained from the Landau Level fan diagram suggests the occurrence of 2D oscillations arising from a Topological Surface State (TSS) for Sb concentrations of x = 0, 0.05 and 0.1; while 2D oscillation seen at higher Sb concentration is attributed to surface 2D electron gas consequent to downward band bending.

---

**Introduction.** – The seminal paper by Zhang et al. [1] on 3D Topological Insulators (TIs) $Bi_2Se_3$, $Bi_2Te_3$ and $Sb_2Te_3$ had pointed out that a similar compound $Sb_2Se_3$ does not show a topologically non-trivial phase. The reason ascribed was the absence of adequate spin-orbit coupling (SOC) strength in $Sb_2Se_3$ that could bring about a band-inversion at the Γ point, observed in the other three materials. It would be of interest to study the effect of substituting Sb with a smaller SOC parameter (of 0.40 eV) in the place of Bi (SOC parameter = 1.25 eV), in $Bi_2Se_3$ and in turn see the effect on the signatures of Topological Surface States in transport measurements viz. Shubnikov de Haas oscillations. Since the end compounds viz., $Bi_2Se_3$ and $Sb_2Se_3$ are topologically distinct; a topological phase transition is expected to occur at a certain critical concentration ($x_c$) of Sb substitution. Independent ab-initio calculations on the $Bi_{2-x}Sb_xSe_3$ system by Liu et al. [2] and Abdalla et al. [3] determined the value of $x_c$ = 1.2, corresponding to the alloy composition $Bi_{0.8}Sb_{1.2}Se_3$. However experimentally, the solid solubility of the $Sb_2Se_3$ in $Bi_2Se_3$ is less than 16 mol% [4] and it is therefore difficult to synthesize single phase crystals with Sb substitution required to observe the phase transition. The transport co-efficients viz. Hall co-efficient $R_H$, electrical conductivity $\sigma$ and the Seebeck co-efficient $\alpha$ of the $Bi_{2-x}Sb_xSe_3$ system as a function of temperature in the solid solubility range were first reported by Drašar et al [5]. Kulbachinskii et al. [6] was the first to observe Shubnikov de Haas oscillations (SdH) on $Bi_{2-x}Sb_xSe_3$ single crystals. However this study was undertaken before $Bi_2Se_3$ was identified as a TI and hence it is worthwhile to repeat the experiment and understand it in present context of the subject of TI. Zhang et al. presented a detailed ARPES study on MBE grown $Bi_{2-x}Sb_xSe_3$ thin films in which they observed that with increase in Sb substitution till x = 0.3, the Dirac point shifted towards the Fermi level and then disappeared for higher concentration [7]. In this experimental report we focus on the understanding the magneto-transport data from single crystals of $Bi_{2-x}Sb_xSe_3$ system.

**Experimental.** – Single crystals of $Bi_{2-x}Sb_xSe_3$ (x = 0.05, 0.1, 0.3, 0.5) were grown by slow cooling stoichiometric melts of high purity Bismuth (Bi), Antimony (Sb) and Selenium (Se) from 850°C to 550°C for over 150 hours (2°C/hour), followed by annealing at 550°C for 24 hours and rapid cooling. Phase elucidation was done by room temperature powder x-ray diffraction (XRD) on powdered single crystals carried out at BL-12 beam-line of Indus II, RRCAT – Indore. The experiment was carried out in transmission geometry wherein x-rays of a defined energy (12.6 keV) were incident on the sample enclosed in a circular depression in kapton tape and the data was collected using a MAR 3450 image plate detector positioned beyond the sample at a distance of 120 mm. The 2D image plate data was converted into a 1D 2θ vs. intensity format using FIT2D software. The Laue diffraction pattern on freshly cleaved crystals was recorded in the transmission mode using a Molybdenum X-ray source and a HD-CR-35 NDT image plate system. Transport measurements were carried out on fresh crystals cleaved from the grown boule via contacts made using 25 micron gold wire and room temperature curing silver paste. Measurement of electrical resistivity as a function of temperature was carried out in a home-made dipper cryostat in the Van der Pauw geometry. Magneto-transport measurements viz. Magneto-resistance [MR] and Hall Effect were carried in a conventional Hall bar geometry in a commercial 15 T cryogen free Magneto-resistance system from Cryogenics Ltd., UK. Positron annihilation spectroscopy measurements were carried out by

(a)Authors to whom any correspondence should be addressed
(a)E -mail: mani@igcar.gov.in; dtr@igcar.gov.in

sandwiching the positron source ($Na^{22}$ evaporated on a 1.25 micron Ni foil) between two crystal surfaces. The positron lifetime was measured using a Fast-Fast coincidence spectrometer having a time resolution of ~260 ps [8] and the data was analysed using LT program [9].

**Results and Discussions.** –

*X-ray diffraction.* Fig. 1(a) shows the powder diffraction data on the $Bi_{2-x}Sb_xSe_3$ crystals with all the peaks indexed to Rhombohedral $Bi_2Se_3$. Absence of peaks corresponding to elemental Sb or Sb-Se secondary phase confirms the substitution of Sb at Bi lattice position. Further a systematic shift in peak positions, in particular (015) peak (shown in Fig. 1(b)) is observed in the powder diffraction pattern, indicating Sb substitution in Bi site. Impurity peaks indexed to elemental Selenium (marked as # in Fig. 1(c)) are observed in the sample with x = 0.5 composition.

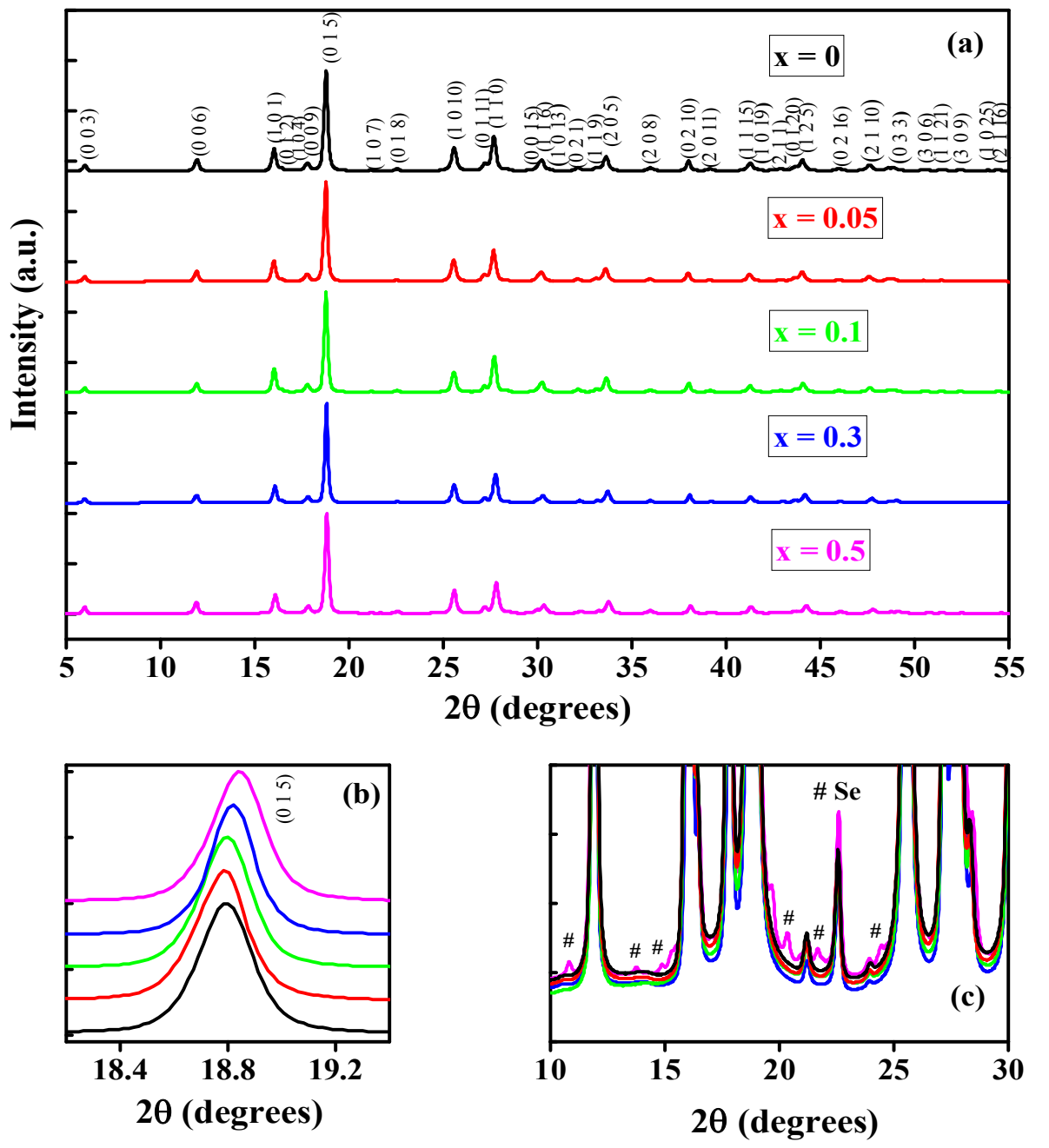

Fig. 1 : (Colour on-line) (a) Powder diffraction pattern of $Bi_{2-x}Sb_xSe_3$ crystals. The peaks have been normalised to the highest observed peak intensity and have been shifted vertically for clarity; (b) observed shift in the 100% peak (015) with increasing Sb substitution; (c) Se impurity peaks observed in sample x = 0.5 denoted by # at their positions.

Fig. 2(a) shows the variation of *a* and *c* lattice parameters as a function of increasing Sb substitution, determined from the powder diffraction patterns using *PowderCell*. It is evident from the figure that there is a non-monotonic variation in the lattice parameters with increasing Sb substitution. In particular, the *a* and *c* lattice parameters for x = 0.05 sample display an anomalous deviation from the overall variation. Fig. 2(b) shows the variation of the unit cell volume as a function of increasing Sb substitution. The decrease of unit cell volume with increasing Sb substitution further substantiates the incorporation of Sb in the lattice, consistent with the smaller ionic radius of Sb when compared to Bi.

We try to reconcile these observations of anomalous *a* and *c* lattice parameters variation with the help of their Laue diffraction patterns shown in Fig. 2(d). It is observed that the Laue spots in the pristine $Bi_2Se_3$ show radial streaking. Such streaking has been attributed to strain inhomogeneities [10][11] along the c-axis arising due the sub-stoichiometry of Se because of inherent Se vacancies in $Bi_2Se_3$ system. On introducing a small amount of Sb (x = 0.05), the lattice stabilizes as witnessed from the sharp spots in the Laue pattern, the c-axis parameter value increases to 28.636 Å. For x = 0.1, the value of c = 28.625 Å and the nature of the Laue spots does not change much from that of x = 0.05. For higher concentrations x = 0.3 and 0.5, the Laue spots become larger and also show slight radial streaking, which can be ascribed to the presence of a uniform strain in the lattice as is evident from the larger values of c = 28.650 Å & 28.669 Å for x = 0.3 & 0.5 respectively. The observed increase in *c* lattice parameter may also be attributed to the reduced cation-anion bonding strength on substitution of the larger and hence more polarisable ion (Bi) with a smaller one (Sb), which could result in an increase in the intra-QL distances and in turn expand the *c* lattice parameter.

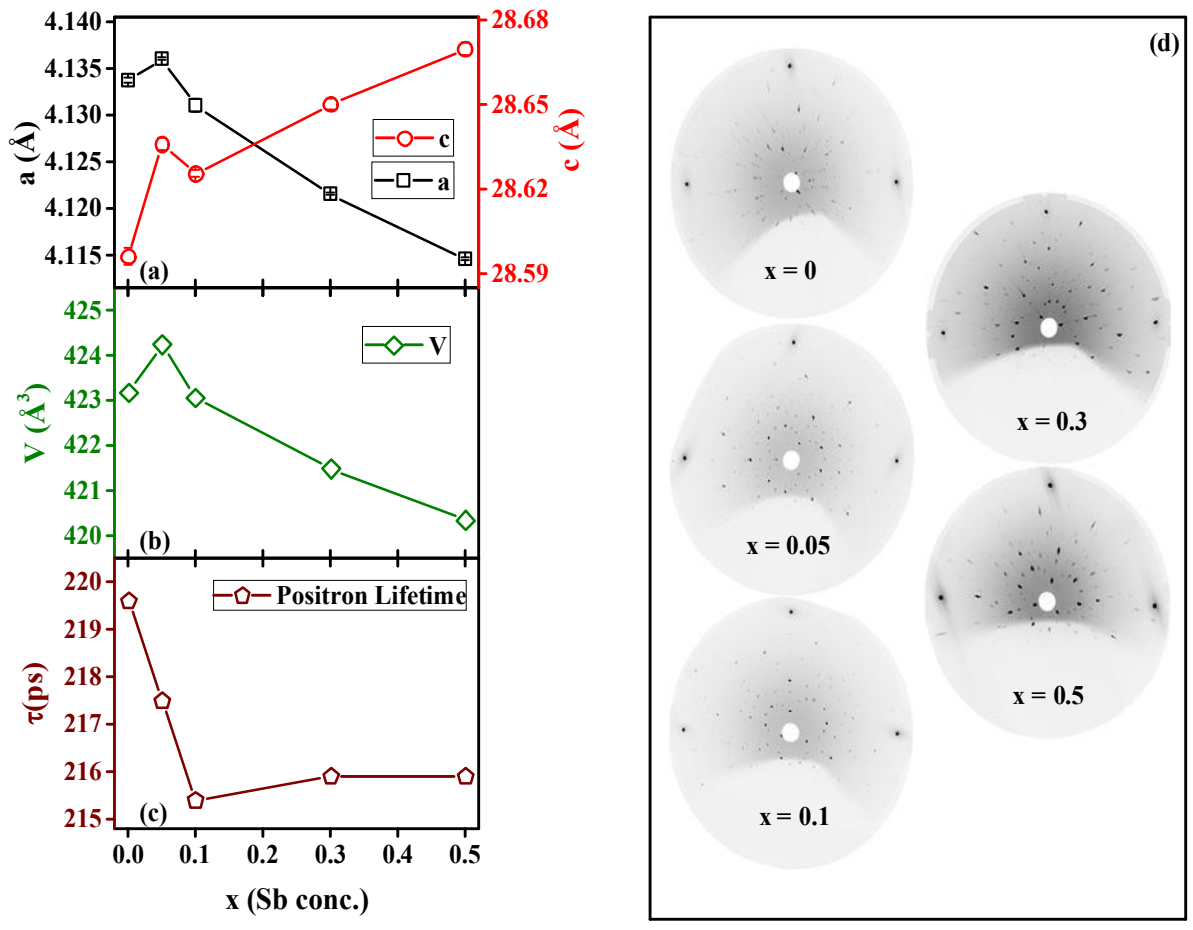

Fig. 2 : (Colour on-line) Variation of (a) *a* and *c* lattice parameters; (b) unit cell volume and (c) Positron Lifetime as a function of Sb concentration in $Bi_{2-x}Sb_xSe_3$; (d) Laue diffraction pattern of $Bi_{2-x}Sb_xSe_3$ crystals in the transmission geometry.

*Resistivity.* Fig. 3(a) shows the temperature dependent resistivity ρ(T) behaviour of the $Bi_{2-x}Sb_xSe_3$ samples. It is evident from the figure that ρ(T) of these samples exhibit metallic behaviour with a positive temperature co-efficient. At low temperatures (T < 30 K) resistivity shows a tendency to saturate. The variation of room temperature resistivity (ρ(300K)) as a function of Sb concentration is shown in inset of Fig. 3(a). ρ(300K) decreases marginally from its value at pristine sample (x = 0) to the low Sb concentration sample x= 0.05, whereas for higher Sb concentrations it shows an increase. The anomalous drop in resistivity value for x = 0.5 sample can be attributed to the presence of the secondary Se phase (cf. Fig. 1(c)). This observed behaviour agrees well with earlier resistivity reports in $Bi_{2-x}Sb_xSe_3$ system [6]. To understand the reason for the observed variation in resistivity with Sb substitution, viz., whether the observed trend is a consequence of reduction in the number of carriers or due to introduction of disorder by Sb substitution, Hall Effect measurements were undertaken.

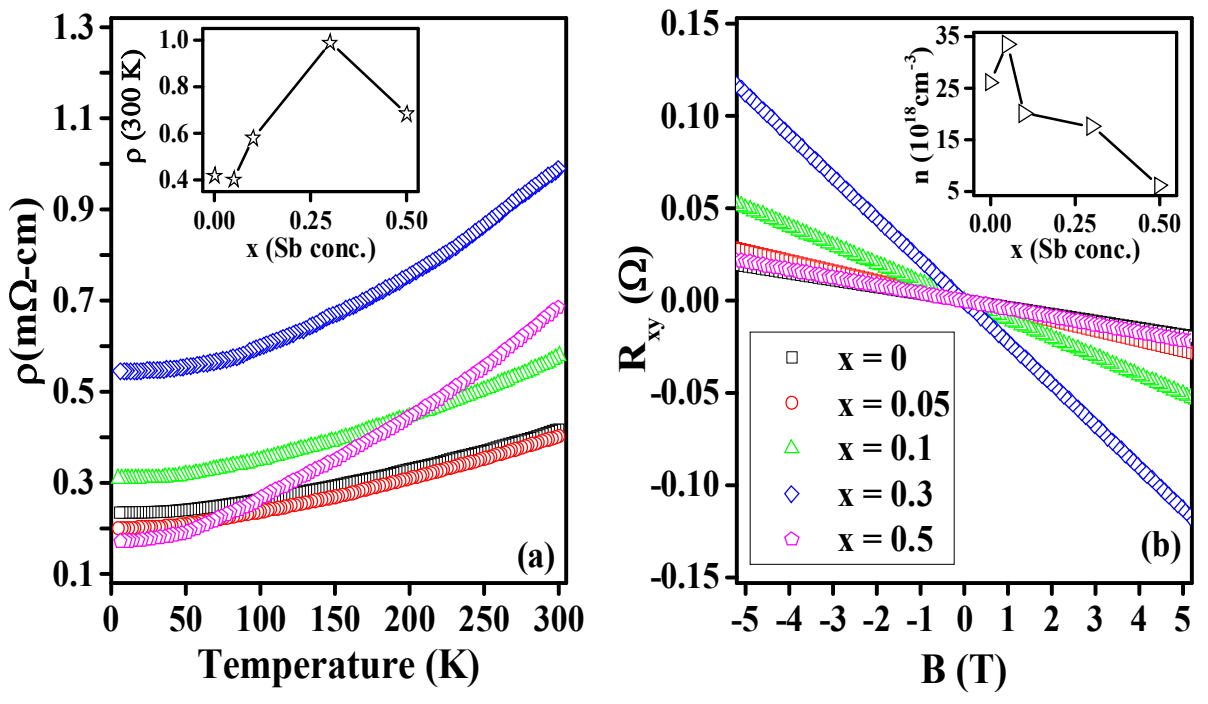

Fig. 3 : (Colour on-line) (a) Resistivity vs. Temperature as a function of Sb concentration, inset shows variation of resistivity as a function Sb concentration at 300 K; (b) Hall resistance ($R_{xy}$) as a function of Sb concentration at 4.2 K, inset shows the charge carrier density as a function of Sb concentration.

*Hall Effect*. Fig. 3(b) shows the Hall Effect data on the various samples at 4.2 K. The linear behaviour and the negative slope indicate that the major charge carriers are electrons. The charge carrier density in $Bi_{2-x}Sb_xSe_3$ system, obtained from the low field slope of the Hall data, is plotted as a function of Sb concentration in inset of Fig. 3(b). It is observed that the carrier density (inset Fig. 3(b)) increases from 26 x $10^{18}$ $cm^{-3}$ for the pristine $Bi_2Se_3$ to 33 x $10^{18}$ $cm^{-3}$ in the low Sb concentration sample x = 0.05. At the higher concentrations of Sb, the charge carrier density (n) decreases monotonically from a value of 33 x $10^{18}$ $cm^{-3}$ for x = 0.05 to 6 x $10^{18}$ $cm^{-3}$ for x = 0.5. The overall variation of charge carrier density with increase in Sb concentration in the pristine $Bi_2Se_3$ lattice is consistent with earlier reports [6]. As Sb is an iso-electronic substituent at the Bi site in $Bi_2Se_3$, no external charges are expected to be introduced in the system. Besides, the decrease in unit cell volume as observed in Fig. 2(b) can only lead to an increase in charge density. This implies that the reduction in charge density seen in the $Bi_{2-x}Sb_xSe_3$ system (ref. inset Fig. 3(b)) is not due to volume change consequent to Sb substitution but could be due to defect introduced into the system by Sb substitution. From our previous study [12] it has been established that selenium vacancies $V_{Se1}$ play an important role in the n-type doping of the bulk $Bi_2Se_3$ system. Thus to look for any possible variation in the $V_{Se1}$ as a function of Sb substitution, we employ positron lifetime measurements.

*Positron Lifetime measurements*. Fig. 2(c) shows the measured positron lifetime values ($\tau$ in ps) in the $Bi_{2-x}Sb_xSe_3$ samples. It is observed that the lifetime values decrease initially, from ~ 220 ps for the pristine $Bi_2Se_3$ to ~ 215 ps for $Bi_{1.9}Sb_{0.1}Se_3$ and then shows no further change in the higher Sb doping. Since the mean positron lifetime τ in a system is given by inverse of the total annihilation rate $\lambda_{tot}$, where $\lambda_{tot}$ is proportional to the enhanced charge carrier density in the system [13]; the observed decrease in τ should imply an increase in charge carrier density if the positron annihilation is purely from the bulk charge density in the $Bi_{2-x}Sb_xSe_3$ samples, which is contrary to the Hall data (cf. Fig. 3(b)). However, it has been brought about in our previous studies [13] that Se1 vacancy in $Bi_2Se_3$ system acts as trap for positron, and hence positron lifetime τ decreases with decrease in concentration of Selenium vacancy $V_{Se1}$ in this system. In this scenario, the observed decrease in the positron lifetime value (see Fig. 2(c)) implies a decrease in concentration of Se vacancy with increase of Sb substitution in $Bi_{2-x}Sb_xSe_3$ upto x=0.1.This also brings consistency in the variation of charge carrier density (n) obtained from Hall measurements and the variation of the positron lifetime as a function of Sb substitution upto x=0.1 (cf. Fig 3(b) & Fig. 2(c)). Both of these results confirm with the assertion that the concentration of $V_{Se1}$ decreases with Sb substitution. Since several calculations [14][15] address the questions of the native defects in $Bi_2Se_3$ based on growth conditions and cell volume, we take stock of literature to examine the possible defects in this system, which can account for the observed decrease in charge carrier density with Sb substitution in $Bi_2Se_3$. As the $Bi_2Se_3$ system is known to have inherent Se vacancies, a crystal grown from stoichiometric precursors tends to be sub-stoichiometric in Se i.e. $Bi_2Se_{3-\delta}$, making it cation rich. Under cation rich conditions, the energies of formation of defects in pristine $Bi_2Se_3$ with increasing order of energies are known to be $V_{Se} \leq Bi_{Se} < Se_{Bi} < V_{Bi}$ [14]. This implies that formation energy for $V_{Se}$ is the lowest among all the defects present in $Bi_2Se_3$ systems. Therefore, reduction in Se(1) vacancies with increase in Sb concentration in $Bi_{2-x}Sb_xSe_3$ samples, can be attributed to the increase in the vacancy formation energy due to reduction in the unit cell volume (as observed in Fig. 2(b)) with increase in x [16]. The positron lifetime beyond x = 0.1 exhibits little change (cf. Fig. 2(c)) which at the first instant may suggest that there is no further decrease in the $V_{Se1}$ vacancy concentration in the crystals with Sb fraction beyond x = 0.1. However, the charge carrier density obtained from Hall Effect measurements shows a further reduction beyond x = 0.1. This implies that carriers other than those arising from $V_{Se1}$, could be responsible for charge density variations seen in samples with x>0.1. The extricate variation of *n* and τ beyond x=0.1 can be understood based on following plausible reasons: Firstly, the observed decrease in *n* could be due to the reduction in the number of $Bi_{Se}$ and $Se_{Bi}$ antisite defects on substitution of Bi by Sb, (as both defects contribute n-type charge carriers in the system ?). The lack of variation in positron lifetime could arise because of positrons not being sensitive to antisite disorder [13]. Secondly, taking clue from an earlier report by Van Vechten [17] which showed that in binary crystals iso-structural to $Bi_2Se_3$, the possibility of formation of a vacancy defect is determined by the surface area of the defect; the larger the surface area, the higher the energy of formation and vice-versa. Since the substitution of Sb at Bi site in $Bi_2Se_3$ leads to a reduction in the effective radius of the cation sub-lattice in $Bi_{2-x}Sb_xSe_3$ system, this may result in a reduction in the formation energy of cation vacancies $V_C$ ($V_{Bi}$), which in turn gives rise to an increase in the number of $V_{Bi}$ defects. Since the $V_{Bi}$ defect level in $Bi_2Se_3$ is known to be a triple acceptor, its net effect would be a reduction in the electron charge density *n*, as seen in Fig. 3(b). On the other hand, as reported in our earlier studies [12], the positron lifetime in uncompensated Bi vacancy is found to be 224 ps, much higher than that in the bulk $Bi_2Se_3$ (201 ps). It means that the positron trapped at $V_{Bi}$ results in an increase in τ as concentration of $V_{Bi}$ increases with increasing x. Therefore, the competing effect of increased in lifetime from positron trapped at $V_{Bi}$ and decrease of τ due to decrease in concentration of $V_{SeI}$ could be responsible for the arresting of decrease in τ at higher Sb concentration x >0.1. The question, why the signature of p-type charge carriers arising from Bi vacancies could not be seen as a non-linear slope of the Hall data versus magnetic field as expected due to the presence of two types of carriers, can be rationalized based on that fact that a triple acceptor with a negligibly small mobility may not show up as a positive contribution in Hall Effect measurement.

*Magneto-resistance.* Fig. 4 shows the Magneto-resistance (MR) plots ($R_{xx}$(B)-$R_{xx}$(0) vs. B) of $Bi_{2-x}Sb_xSe_3$ samples at 4.2 K in two different configurations – field out-of-plane (H||c) and field in-plane (H||ab). $R_{xx}$ is the longitudinal resistance at each field applied and $R_{xx}$(0) is the value of resistance at 0 T. It is observed that the Shubnikov de-Haas (SdH) oscillations in MR occur only in H||c configuration for the samples with x = 0, 0.05, 0.1 & 0.3, while for sample with x = 0.5 the SdH oscillations are seen in both H||c and H||ab configurations.

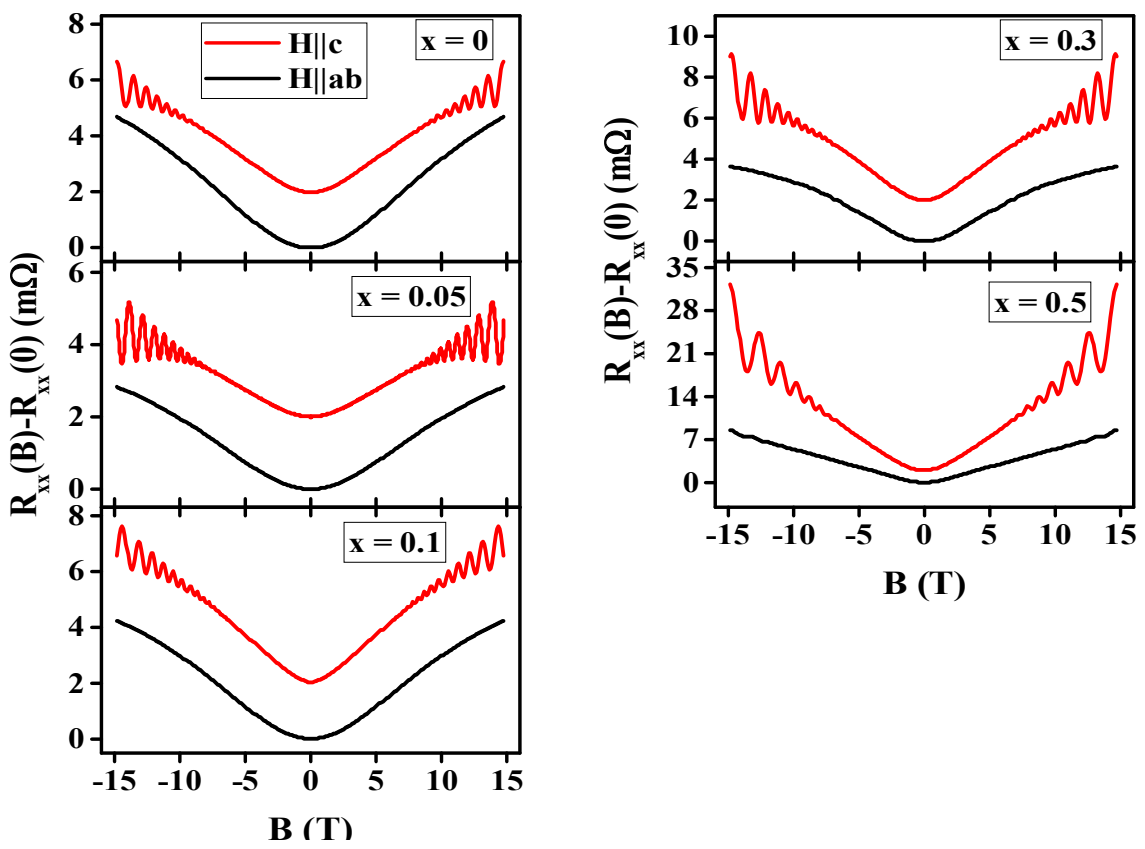

Fig. 4 : (Colour on-line) Magneto-resistance data of $Bi_{2-x}Sb_xSe_3$ crystals at 4.2 K. The data in both H||c and H||ab configuration are plotted in the same panel for each composition and are offset for clarity

It is known that in angular dependent magneto-resistance measurements, the positions of maxima and minima in oscillations depend only on the perpendicular component of field ($B_\perp$) as $(B_\perp)^{-1} \equiv (B\cos\theta)^{-1}$, where θ is the angle between the field direction and the normal to the crystal surface [18]. If the oscillations disappear for θ = 90° i.e. the H||ab (field in-plane) configuration, the SdH oscillations could be from a 2D Fermi surface. It is evident from Fig. 4 that oscillations are not observed for H||ab configuration for x = 0, 0.05, 0.1 and 0.3 crystals. This clearly indicates that the observed SdH oscillations for these compositions arise from an underlying 2D Fermi Surface. In x = 0.5 sample, oscillations are observed in both H||c and H||ab plane which suggest a bulk 3D origin. The SdH oscillations obtained experimentally are further analysed using the Lifshitz-Kosevich (L-K) equation [19] after suitable background subtraction. This fit results in the extraction of valuable parameters such as Fermi wave-vector ($k_F$), Fermi velocity ($v_F$). The L-K equation [19] is given as $\Delta R_{xx} = A_0 R_T R_D R_S \cos\left[2\pi\left(\frac{F}{B}+\frac{1}{2}-\beta\right)\right]$, where $\Delta R_{xx}$ = oscillatory component of resistance after subtracting the background obtained from the experiments shown in Fig. 4; F = Frequency of oscillations obtained from a Fourier transform on the oscillatory component $\Delta R_{xx}$, B = external magnetic field in T, $\beta$ = Berry phase γ divided by 2π; which is 0.5 for Dirac fermions possessing linear energy dispersion and 0 for fermions with parabolic dispersion the pre-factors; $A_0 = \left(\frac{\hbar\omega_c}{2\varepsilon_F}\right)^{\frac{1}{2}}$ = constant; $R_T = \frac{\left(2\pi^2 k_B T/\hbar\omega_c\right)}{\sinh\left[2\pi^2 k_B T/\hbar\omega_c\right]}$ = Temperature damping factor; $R_D = e^{-\left[\frac{2\pi^2 k_B T_D}{\hbar\omega_c}\right]}$ = Dingle damping factor; $R_S = \cos\left(\frac{1}{2}\pi g\frac{m_0}{m^*}\right)$ = Spin damping factor. Here $\varepsilon_F$ = Fermi energy, $\omega_c$ = cyclotron frequency, $k_B$ = Boltzmann constant, $T_D$ = Dingle temperature, g = gyromagnetic ratio, m* = effective mass in terms of electron rest mass $m_0$. Fig. 5(a) shows a representative fit of the L-K equation to the oscillatory component $\Delta R_{xx}$ for the case of pristine $Bi_2Se_3$. The fitting parameters $k_F$, $T_D$, $m_0$, β, g, quantum scattering time $[\tau_q = \hbar/(2\pi k_B T_D)]$, Fermi velocity$[v_F = (\hbar k_F)/m^*]$, 2D carrier density$[n_s = (k_F^2)/4\pi]$, Surface mean free path $[l_S^{SdH} = v_F\tau_q]$ and surface mobility $[\mu_S^{SdH} = (e l_S^{SdH})/(\hbar k_F)]$ obtained from the L-K equation fit to the MR data of $Bi_{2-x}Sb_xSe_3$ samples (x = 0, 0.05, 0.1, 0.3, 0.5) are tabulated in Table 1. The value of oscillation frequency in the H||c direction and some of the fit parameters and their variation with Sb concentration are shown in Fig. 6. The β value for x = 0.5 sample is not considered as the oscillations exhibit 3D character and thus are likely to arise from the quantization of the bulk 3D Fermi surface of $Bi_2Se_3$ as observed by Eto et al [20].

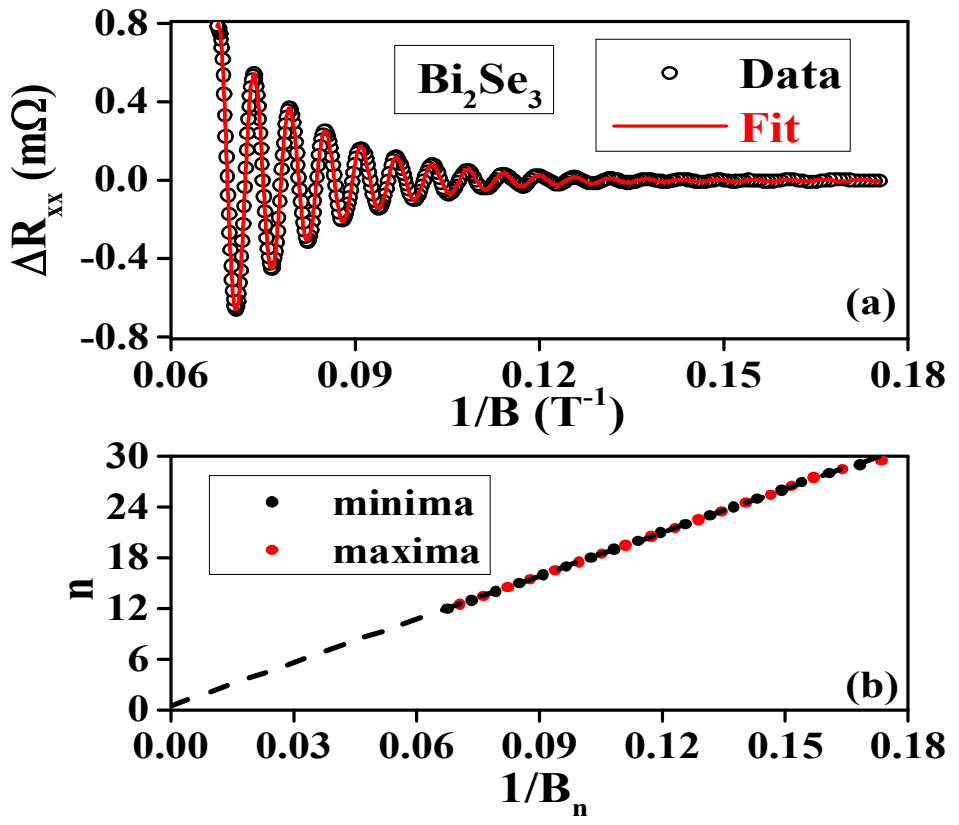

Fig. 5 : (Colour on-line) (a) Representative fit of L-K equation fit to the SdH oscillations in $Bi_2Se_3$ sample; (b) LL Fan diagram analysis

The oscillation frequency in x = 0 sample is 172 T, which increases marginally to 174 T in x = 0.05 sample. For higher Sb concentrations viz. x = 0.1, 0.3 and 0.5 the frequency decreases. The Fermi wave-vector ($k_F$), defined as the radius of the 2D Fermi sphere, shows a marginal increase in value from the pristine x = 0 to x = 0.05 sample and then decreases for higher concentrations of Sb substitution. The reduction in its value implies a reduction in the cross-section of the Fermi level at $E_F$. The cyclotron mass of 0.152 $m_0$ observed in x = 0 reduces to 0.117 for x = 0.05 and nearly remains the same for the higher Sb substitutions. The values of Fermi-velocity ($v_F$) obtained from the values of $k_F$ and $m_0$ shows a variation with Sb, which is similar to that seen for $k_F$. The magnitude of the 2D carrier density $n_s$ shows a decreasing trend with increasing Sb substitution. The dingle temperature ($T_D$), quantum scattering time ($\tau_q$) and surface mean free path ($l_s^{SdH}$) do not show any systematic variation with Sb substitution implying that Sb substitution in $Bi_2Se_3$ does not introduce any disorder into the system. These results are consistent with earlier calculations on the $Bi_{2-x}Sb_xSe_3$ system [2].

respectively and plotted vs. $1/B_n$. The value of intercept at 0 $T^{-1}$, obtained from a linear fit to the data; with slope being fixed by the frequency of oscillations helps in determining the exact nature of the 2D oscillations. For TSS, the presence of a $\pi$ Berry phase should have an intercept of 0.5 in the LL fan diagram. A representative LL-Fan diagram for the x = 0 i.e. $Bi_2Se_3$ sample is shown in Fig. 5(b). The values of intercepts obtained from the LL fan diagram analysis for the compositions x = 0, 0.05, 0.1 and 0.3 are plotted in Fig. 6.The intercept value changes from 0.5 in pristine to 0.6 in x = 0.05 and x = 0.1. For x = 0.3 sample the intercept value further deviates to 0.875. The value of β obtained from the L-K equation fit to SdH oscillations are also plotted simultaneously in Fig. 6.

Table 1. Fitting parameters and derived physical constants from L-K equation fit to SdH oscillations in $Bi_{2-x}Sb_xSe_3$ system.

| x in $Bi_{2-x}Sb_xSe_3$ | $k_F$ ($10^6$ $cm^{-1}$) | $T_D$ (K) | $m^*$ ($m_0$) | $\tau_q$ (fs) | $v_F$ ($10^5$ $ms^{-1}$) | $n_s$ ($10^{12}$ $cm^{-2}$) | $l_s^{SdH}$ (nm) | $\mu_s^{SdH}$ ($cm^2$/V.s) | β | g |
|---|---|---|---|---|---|---|---|---|---|---|
| 0 | 7.23 | 25 | 0.152 | 47 | 5.49 | 4.16 | 25 | 541 | 0.5 | 2 |
| 0.05 | 7.27 | 27 | 0.117 | 44 | 7.19 | 4.21 | 31 | 659 | 0.6 | 2 |
| 0.1 | 6.85 | 22 | 0.117 | 54 | 6.75 | 3.73 | 35 | 804 | 0.6 | 2 |
| 0.3 | 6.54 | 26 | 0.117 | 46 | 6.47 | 3.40 | 29 | 690 | 0.875 | 2 |
| 0.5 | 5.08 | 23 | 0.116 | 52 | 5.06 | 2.05 | 26 | 785 | - | 2 |

The surface mobility ($\mu_s^{SdH}$) initially increases from 541 $cm^2$/V.s in x = 0 to 804 $cm^2$/V.s in x = 0.1, then reduces to 690 $cm^2$/V.s for x = 0.3 sample and increases again to 785 $cm^2$/V.s in x = 0.5. The value of β for x= 0, 0.05, 0.1, 0.3 are 0.5, 0.56, 0.7, 0.875 respectively. The value of g obtained is 2 for all compositions.

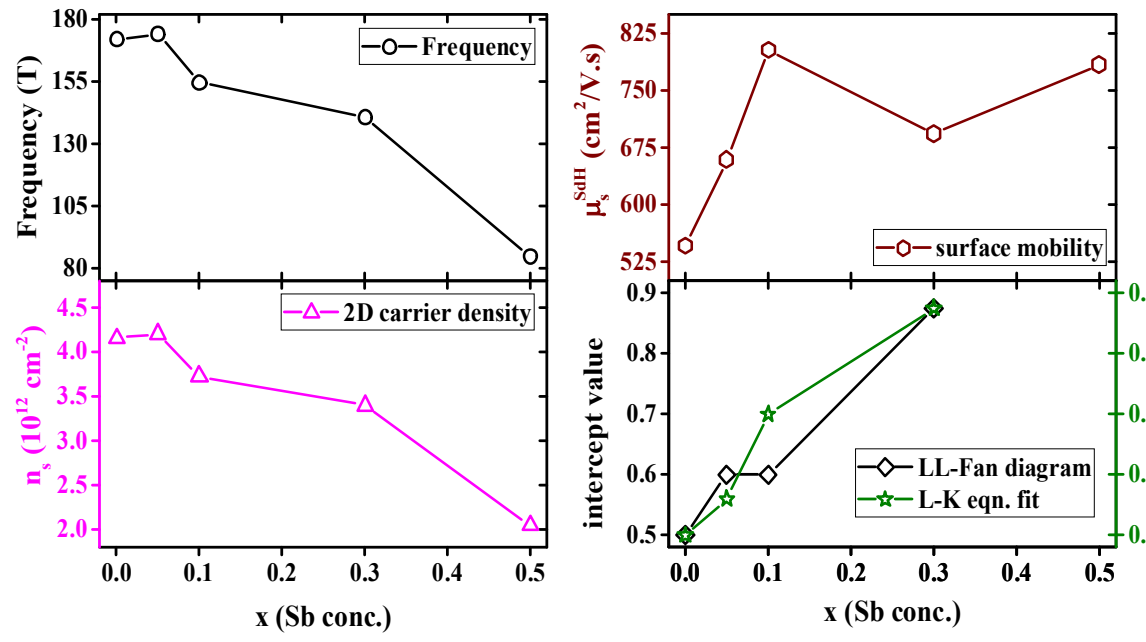

Fig. 6 : (Colour on-line) Variation of Frequency (F), 2D carrier concentration ($n_s$), surface mobility ($\mu_s^{SdH}$), β and the LL-fan diagram intercept as a function of Sb concentration in $Bi_{2-x}Sb_xSe_3$ single crystals.

Having observed experimentally that the SdH oscillations in $Bi_{2-x}Sb_xSe_3$ have a 2D origin, it is important to ascertain whether these oscillations arise solely from the relativistic Dirac fermions of the TSS that have linear dispersion or from a trivial 2D electron gas (2DEG) that forms at the interface of semiconducting bulk/metallic surface as a consequence of equilibration between the bulk and surface Fermi levels and have a parabolic dispersion [21]. This is done with the help of the Landau Level (LL) fan diagram analysis ($1/B_n$ vs. n) of the SdH oscillations. To plot LL fan diagram in TI systems, the oscillations in resistivity are first converted to oscillations in conductivity using the relation $\sigma_{xx} = \rho_{xx}/(\rho_{xx}^2 + \rho_{xy}^2)$ [22] following which the minima and maxima in conductivity oscillations are assigned integer and half-integer values

*Discussion*. The Berry phase of 0.5, obtained from the LL-fan diagram analysis and a Fermi velocity value of $v_F \sim 5.49$ x $10^5$ $ms^{-1}$ obtained from the L-K equation fit to the SdH oscillations which is in agreement with the Fermi velocity value of 5 x $10^5$ $ms^{-1}$ for the Dirac cone states in $Bi_2Se_3$ as observed from ARPES measurements reported earlier [23] gives strong evidence for the presence of TSS signatures in transport in the pristine x = 0 sample. In addition the single valued frequency value of 172 T corresponds to a large Fermi surface as expected for TSS. In the $Bi_2Se_3$ system, SdH oscillations from the surface states can be realized in TIs only when there is an upward band bending in the system where the surface Fermi level lies below the bulk Fermi level in energy [21]. With the origin of oscillation in the pristine sample assigned to the TSS using Berry phase values from the LL fan diagram analysis (Fig. 5) and $v_F$ (Table 1.), it is clear that the band-bending in our pristine $Bi_2Se_3$ system is upward. The 2D carrier density of $n_s \sim 4.12$ x $10^{12}$ $cm^{-2}$, estimated from the SdH analysis is also well within the theoretically calculated value of 5 x $10^{12}$ $cm^{-2}$, below which upward band bending is expected in pristine $Bi_2Se_3$ [21]. The introduction of Sb into the lattice is seen to give rise to two visible changes in the MR data viz. the change in frequency of oscillations (ref. Fig. 6) and the change in LL Fan diagram intercept (ref. Fig. 6). The reduction in frequency of oscillations could arise due to a reduction in the Fermi level, as a consequence of an upward shift of the Dirac Point (DP) due to reduction in the bulk band-gap. Such an upward shift of the DP in Sb substituted $Bi_2Se_3$ system has been predicted from band-structure calculations [2][3] and also been observed by ARPES in thin films of Sb substituted $Bi_2Se_3$ films grown by MBE technique [7]. The deviation in LL-Fan diagram intercept is marginal; from 0.5 for x = 0 to 0.6 for x = 0.05 & 0.1 samples, well within the spread observed in literature [18] and thus the SdH oscillations observed in x = 0.05 & 0.1 can still be considered as the signature of TSS states. For x = 0.3, the intercept deviates significantly to a value of 0.875. This could imply that the observed SdH oscillations cannot be purely attributed to the TSS. Alternatively, the 2D oscillations in x = 0.3 sample could arise from a trivial 2D state of 2DEG. This could arise due to the expected changes in Band gap with Sb substitution [2][3]. Since the band-gap is expected to reduce with Sb substitution, as mentioned earlier the DP shifts to higher energy with increase in Sb substitution [7]. It can then be conceived that the shift could result in the swapping of the Fermi level positions, such that in $E_F^{surface}$>$E_F^{bulk}$leading to surface charge accumulation and hence formation of Quantum Well States (QWS). This QWS at the surface being inherently 2D in nature, can show SdH oscillations of 2D character, but have a parabolic dispersion such that their Berry phase deviates from 0.5. Thus the variation in the L-K equation parameters could then be reconciled with transport arising from TSS moving over to a topologically trivial 2DEG state. This cross-over made plausible with Sb substitution is illustrated schematically in Fig. 7.

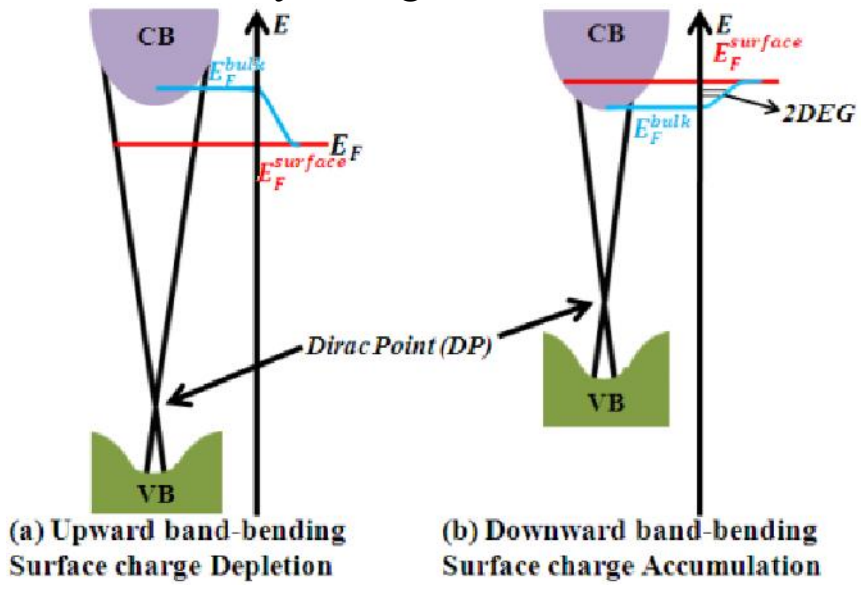

Fig. 7 : (Colour on-line) Schematic diagram of Band structure in $Bi_{2-x}Sb_xSe_3$ showing (a) upward band bending and (b) downward band-bending and also reduction of bulk band gap [2][3] with Sb substitution. $E_F^{Bulk}$ and $E_F^{Surface}$are the bulk Fermi level and surface Fermi level respectively. DP corresponds to the Dirac Point and 2DEG corresponds to the trivial 2D electron gas that can form due to surface charge accumulation. The bulk Fermi level is pinned to the bottom of the conduction band as defined by the value of carrier concentration obtained from Hall Effect measurement [21]

*Summary and Conclusion*. Synchrotron measurements suggests that the $Bi_{2-x}Sb_xSe_3$ (x = 0, 0.05, 0.1, 0.3, 0.5) crystal system is highly single phase only for low Sb substitution concentration i.e. ≤ 15%. The resistivity increases with increase in Sb substitution which is supported by the observed reduction in charge carrier density from Hall Effect measurements. From analysis of the SdH oscillations obtained from magneto-resistance measurements we surmise that the signatures of the TSS are observed for Sb concentration of upto x = 0.1, whereas for x = 0.3 the trivial 2DEG masks the transport from TSS. Thus we can conclude that a reduction in the bulk band-gap can shift the DP in such a manner that the positions of the Fermi levels interchange leading to a transport being dominated by the electronic structure of the bulk/surface interface which can mask signatures from TSS in transport measurement.

***

**Acknowledgements.** One of the authors, T. R. Devidas gratefully acknowledges the Department of Atomic Energy for providing the Senior Research Fellowship and Dr. A. K. Sinha, RRCAT - Indore for the powder x-ray diffraction data on the various samples. The authors gratefully acknowledge UGC-DAE-CSR node at Kalpakkam for providing access to the 15 T cryogen free magneto-resistance facility. Prof. A. Thamizhavel (TIFR, Mumbai) is thanked for the Laue diffraction patterns on all the samples.